\documentclass{aa}
\usepackage{natbib,amssymb}
\usepackage[pdftex]{graphicx,color}
\usepackage{rotating,footmisc}
\bibpunct{(}{)}{;}{a}{}{,}
\usepackage[pdftex]{graphicx,color} 
\usepackage{txfonts}
\begin{document}
   \title{Integral field optical spectroscopy of a representative sample of ULIRGs:}

   \subtitle{I. The Data\thanks{Based on observations with the William Herschel Telescope
  operated on the island of La Palma by the ING in the Spanish Observatorio
  del Roque de los Muchachos of the Instituto de Astrof\'{\i}sica de
  Canarias. Based also on observations with the NASA-ESA Hubble Space
  Telescope, obtained at the Space Telescope and Science Institute, which is
  operated by the Association of Universities for Research in Astronomy,
  Inc. under NASA contract number NAS5-26555.}}

   \author{M. Garc\'{\i}a-Mar\'{\i}n
          \inst{1, 2}
          \and
          L. Colina\inst{1}
          \and
          S. Arribas\inst{1}
          \and
          A. Monreal-Ibero\inst{3}
          }
\offprints{M. Garc\'{\i}a-Mar\'{\i}n (maca@ph1.uni-koeln.de)}
   \institute{Departamento de Astrof\'{\i}sica Molecular e Infrarroja, Instituto de Estructura de la Materia, CSIC
              C/ Serrano 121, Madrid, Spain\\
         \and
             Present Address: I. Physikalisches Institut, Universit\"at zu K\"oln
Z\"ulpicher Strasse 77, 50937 K\"oln, Germany\\
             \email{maca@ph1.uni-koeln.de}
          \and
             European Southern Observatory, Karl-Schwarzschild-Strasse 2
        D-85748 Garching bei M\"unchen, Germany}

   \date{Received ; accepted}

 
  \abstract
   {Ultraluminous infrared galaxies are among the brightest objects in the local Universe. They are powered by strong star formation and/or an AGN. They are also likely to be the progenitors of elliptical galaxies. The study of the structure and kinematics of samples of local ULIRGs is necessary to understand the physical processes that these galaxies undergo, and their implications for our understanding of similar types of galaxies at high redshift.
}
   {The goal of the project is to analyze the structure, dust distribution, ionization state, and kinematics of a representative sample of 22 ULIRGs. The galaxies in the sample undergo different merger phases (they are evenly divided between pre- and post-coalescence systems) and ionization stages (27\% H\,{\sc ii}, 32\% LINER, 18\% Seyfert, and 23\% mixed classifications) over a wide infrared luminosity range (11.8$\leq$L$_{IR}$/L$_{\odot}$$\leq$12.6), with some galaxies of low-luminosity. The main aims of this paper are to present the sample and discuss the structure of the stellar and ionized gas components.
}
   {Our study relies on the use of integral field optical spectroscopy data obtained with the INTEGRAL instrument at the William Herschel Telescope.  
}
   {The structure of the ionized gas as traced by different emission lines has been studied and compared with that of the stellar continuum. We find structural variations between the gaseous and the stellar components, with offsets in the emission peaks positions of up to about 8~kpc. Young star formation (as traced by the H$\alpha$ emission) is present in all regions of the galaxies. However, for 64\% of ULIRGs in an early interaction phase, the young star formation peak does not coincide with the stellar maxima. In contrast, galaxies undergoing advanced mergers have a H$\alpha$ peak that is located in their nuclear regions. In three of the studied ULIRGs, hard ionizing photons traced by the [O\,{\sc iii}]$\lambda$5007 line excite extra-nuclear nebulae out to distances of about 7~kpc. These regions do not show bright stellar emission, but are rather dominated by nebular emission. These galaxies have nuclei classified as Seyfert in the literature. Approximately 40\% of the pre-coalescence ULIRGs exhibit shifts between the peaks of their red continuum and that local to the [O\,{\sc i}]$\lambda$6300 line. However, some of these peaks are associated with the secondary stellar nucleus of the system. In contrast, the emission in post-coalescence ULIRGs is concentrated towards the nuclei. These results imply that evolution caused by a merger is ocurring in the ionized gas structure of ULIRGs.}
   {}

   \titlerunning{IFS of a Representative Sample of ULIRGs. I. The Data}
   \authorrunning{Garc\'{\i}a-Mar\'{\i}n et al.}  

   \maketitle
%

\section{Introduction}

Ultraluminous Infrared Galaxies (ULIRGs, 10$^{12}$L$_{\odot}$$\le$ L$_{bol}$$\sim$ L$_{IR}$$[8-1000\mu m]$ $\le$ 10$^{13}$L$_{\odot}$) are a key galaxy class characterized by the emission of the bulk of their energy in the infrared. They are as luminous as QSOs and, in the local Universe, twice as numerous (see reviews by Sanders \& Mirabel 1996; Lonsdale et al.\,2006). The existence of galaxies with a strong infrared excess was reported by Low \& Kleinmann (1968) and Rieke \& Low (1972), but the study of ULIRGs as a galaxy class started with their detection in large numbers by the IRAS satellite (e.g., Soifer et al.\,1984; Sanders et al.\,1988). Subsequently, optical long-slit spectroscopy was used to investigate their main energy source (e.g., Veilleux et al.\,1995). It has been generally accepted that ULIRGs are powered mainly by intense star formation, whereas the presence and relative importance of an AGN was a matter of debate (Genzel et al.\,1998; Veilleux, Kim \& Sanders 1999; Risaliti et al. 2006; Farrah et al. 2007). However, IR studies suggest that the contribution of AGN to the bolometric luminosity of the system may be relevant in 15--20\% of cases (see Nardini et al.\,2008 and Risaliti et al.\,2006).

ULIRGs are characterized by interaction/merger processes that strongly drive their dynamics. The merger triggers the starburst activity (i.e.,the ultra-luminous phase), is responsible for the complex tidally-dominated morphologies (e.g., Borne et al.\,2000; Bushouse et al.\,2002; Farrah et al.\,2001), dominates the ionized gas kinematics (e.g., Colina, Arribas \& Monreal 2005), and possibly transforms spiral galaxies into intermediate mass ellipticals (e.g., Genzel et al.\,2001; Tacconi et al.\,2002; Dasyra et al.\,2006). 

In a cosmological context, the importance of ULIRGs is significant because they appear to be the low-{\textit{z}} analogs to at least part of the population of star-forming luminous galaxies giving rise to the far-IR background. These are the sub-mm galaxies (Smail, Ivison \& Blain 1997) and the ULIRGs detected by \textit{Spitzer} (P\'erez-Gonz\'alez et al.\,2005). It has been shown that, in contrast to the situation in the local Universe, the importance of the contribution of ULIRGs and the less luminous LIRGs (Luminous Infrared Galaxies, 10$^{11}$L$_{\odot}$$\le$ L$_{bol}$$\sim$ L$_{IR}$$[8-1000\mu m]$ $\le$ 10$^{12}$L$_{\odot}$) to the IR extragalactic light increases with redshift (Lagache, Puget \& Dole 2005; Le Floc'h et al.\,2004; Lonsdale et al.\,2004; Yan et al.\,2004; Caputi et al.\,2006a, 2006b). They are the main contributors to the comoving star density of the Universe at z$>$1.0 (Elbaz et al.\,2002; Le Floc'h et al.\,2005; P\'erez-Gonz\'alez et al.\,2005; Caputi et al.\,2007). Generally, these high redshift studies do not focus on the detailed characterization of individual objects, but rather on their general integrated properties. To be able to study spatially resolved information, it is however important to understand their internal structure, dust distribution, ionization structure, and kinematics. This information is limited at high redshifts, and studies have used integral field spectrographs to analyze these high redshift objects in detail (e.g., Genzel et al\,2006; F\"orster-Schreiber et al.\,2006, 2009). 

In the local Universe, where the spatial resolution allows us to study the internal galaxy structure in greater detail, comprehensive studies of local ULIRGs have been completed using imaging and long-slit spectroscopy in the optical and infrared (e.g., Veilleux et al.\,1995; Genzel et al.\,1998; Kim et al.\,1998; Scoville et al.\,2000; Farrah et al.\,2001; Bushouse et al.\,2002; Imanishi et al.\,2007). In some cases, the combination of these images with optical two-dimensional spectral information such as that provided by integral field spectroscopy (IFS) has been used to carry out comprehensive studies of individual or a few ULIRGs (e.g., Arribas, Colina \& Borne 1999; Monreal-Ibero 2004; Colina, Arribas \& Monreal-Ibero 2005). 

To obtain statistical knowledge of the nature of ULIRGs, we have started a program to study, on the kpc scale, the dust structure, the two-dimensional ionization statem, and the ionized gas kinematics of a representative sample of local ULIRGs, based on optical IFS data obtained with the instrument INTEGRAL. A similar study but for a representative sample of LIRGs is being conducted using VIMOS at the VLT (Arribas et al.\,2008) and PMAS on the Calar Alto 3.5~m (Alonso-Herrero et al.\,2009). All these studies are part of the same project, which investigates the two-dimensional extinction, ionization, and kinematic kpc-scale structure of a representative sample of low$-z$ LIRGs and ULIRGs, along with its implications for their high-redshift analogs.  For instance, the IFS data presented in this paper sample regions similar to those that will be studied by the instruments MIRI (Mid IR Instrument) and NIRSpec (Near Infrared Spectrograph), due to be launched onboard of the \textit{James Webb Space Telescope}, at redshifts above 1.

In this paper, we combine data from previously published works (see Colina, Arribas \& Monreal 2005; Garc\'{\i}a-Mar\'{\i}n et al.\,2006 and references therein) and new data of 13 unpublished galaxies. The paper is organized as follows. We describe the sample selection in Sect.\,2; the observations and data reduction are outlined in Sect.\,3, while the data analysis is presented in Sect.\,4. The morphology of the continuum stellar emission and ionized gas is presented in Sect.\,5. Finally, the summary is given in Sect.\,6. The second paper of the series will present the extinction of the sample. The third paper will focus on the ionization state, whereas the fourth will present the kinematics. Throughout the paper, we use $\Omega_{\Lambda}$=0.7, $\Omega_{M}$=0.3, and H$_{0}$=70 km s$^{-1}$ Mpc$^{-1}$.


\section{Sample selection and properties}

\begin{figure}[t]
   \begin{minipage}{\columnwidth}
   \centering
   \includegraphics[width=\columnwidth]{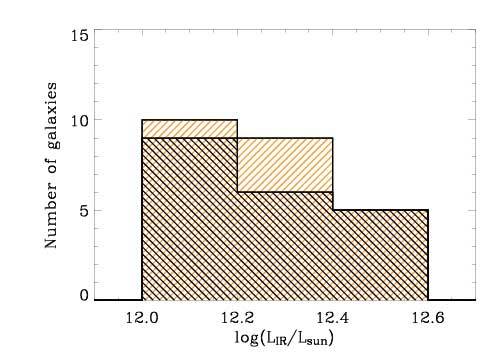}
      \caption{Luminosity distribution of the IRAS Revised BGS (Sanders et al.\,2003, orange dashed histogram) compared with our sample of ULIRGs (black dashed histogram). The luminosities of the BGS galaxies have been re-calculated to consistently compare both datasets. The luminosity range is 12.0$\le$log(L$_{IR}$/L$_{\odot}$) $\le$12.6 with bins of 0.2. The two LIRGs of the sample have not been included in the graphic.}
         \label{fig0}
   \end{minipage}
   \end{figure}

The present sample of low-\textit{z} ULIRGs was selected using the following criteria: (1) to sample the IR luminosity range (Fig. \ref{fig0}); (2) to cover all types of nuclear activity, that is different excitation mechanisms such as H\,{\sc{ii}}- (i.e., star formation activity), LINER- (i.e., superwinds, shocks), and Seyfert-like (i.e., presence of an AGN); (3) to optimize the linear scales by selecting low-\textit{z} galaxies; and (4) to span different phases of the interaction process.

\begin{figure*}
\centering
\includegraphics[width=\textwidth]{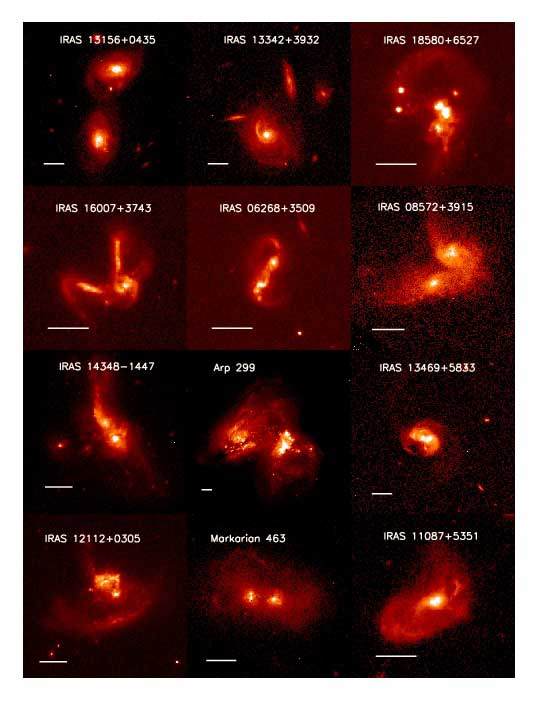}
\caption{\textit{HST} (WFPC2/F814W) images of our sample of galaxies. From left to right, up to down, the projected nuclear separation decreases (i.e., the merger evolves from pre- to post-coalescence galaxies). Orientation is north up east to the left. The scale indicates 5 arcsec.}\label{galaxias1WFPC2}
\end{figure*}

We selected our sample from those Sanders et al.\,(1988; 1995), Melnick \& Mirabel (1990), Leech et al. (1994), Kim et al. (1995), Lawrence et al. (1996), and Clements et al. (1996). Our final selection of 22 northern hemisphere ULIRGs (33 individual galaxies) is shown in Fig. \ref{galaxias1WFPC2}, and its main characteristics are presented in Table\,\ref{TableSample}. In the following paragraphs, we describe the selection criteria in detail.

The range of luminosities covered by the sample is 11.8$\le$log(L$_{IR}$/L$_{\odot}$) $\le$12.6, including two systems classified as LIRGs, but close enough to the low IR luminosity range of the ULIRG class. The luminosity distribution of the galaxies in our sample is similar to that of the complete flux-limited\footnote{threshold criteria S$_{60\mu \rm{m}}$$>$5.4 Jy, referred to the IRAS band centered in 60$\mu$m} IRAS Revised Bright Galaxy Sample (RBGS; Sanders et al.\,2003), which covers the entire sky surveyed by IRAS at Galactic latitudes $|$b$|$$>$5$^{\circ}$ (Fig. \ref{fig0}).
%

 All the excitation mechanisms are well represented (see Table\,\ref{TableSample}): 27\%\,, 32\%\,, and 18\%\,of the ULIRGs are classified as H\,{\sc ii}, LINER, and Seyfert, respectively. The remaining 23\%\,represent objects with unclear or mixed classifications. Using the 1 Jy sample, a flux-limited sample of ULIRGs identified from the IRAS Faint Source Catalog, Kim et al. (1998) derived the following distribution for ULIRGs: 28\%\,for H\,{\sc ii}, 38\%\,for LINER, and 34\%\,for Seyfert. The H\,{\sc ii} and LINER fractions compare well with ours, whereas the Seyfert fraction is almost double. This difference is mainly caused by the wider luminosity covered by Kim et al. (1998; their sample goes up to L$_{IR}$=12.8L$_\odot$, whereas ours stops at 12.6L$_\odot$), which results in a higher percentage of AGN (Veilleux et al.\,1999). 

The galaxies are located at distances of between about 40 and 900 Mpc, with an angular sampling that ranges from 0.2 to 3.1 kpc arcsec$^{-1}$ (see individual values in Table\,\ref{TableSample}). The median redshift is 0.11, which compares well with 0.14, the median value of the 1 Jy sample (Kim \& Sanders 1998). 

Finally, the ULIRGs in our sample exhibit a variety of morphologies, from interacting disks with a projected nuclear distance up to 36~kpc, to close nuclei or individual nucleus surrounded by a common stellar envelope. There are detailed classification schemes based on the interaction stage and morphological features that lead to the existence of many classes and subclasses (see e.g., Surace 1998; Veilleux et al.\,2002). We, however, chose to separate our sample of ULIRGs into two broad categories using the projected nuclear distance as a discriminator: (1) \textit{Pre-coalescence galaxies:} those with a projected nuclear separation larger than 1.5~kpc. (2) \textit{Post-coalescence galaxies:} those with a projected nuclear separation smaller or similar than 1.5 kpc. This accurate distance calculation was possible because of the high spatial resolution of the \textit{HST} images (typical scale 0\farcs1 per pixel). The selected distance evenly distributes the galaxies into the two categories, and is appropriate for separating ULIRGs into early and late merger phases. In addition to this, theoretical models suggest that by the time the two nuclei have reached a separation of $\lesssim$ 1~kpc, the stellar system has basically achieved equilibrium, although their nuclei can still be separate structures (e.g., Mihos 1999; Mihos \& Hernquist 1996; Bendo \& Barnes 2000; Naab et al.\,2006). However, it is important to clearly state that the use of projected nuclear distances may lead to some misclassification because galaxies identified as post-coalescence galaxies could well be in a pre-coalescence state. 

Considering all these aspects, we conclude that our sample is essentially representative of the ULIRG class.

\addtocounter{figure}{-1}
\begin{figure*}
\centering
\includegraphics[width=\textwidth]{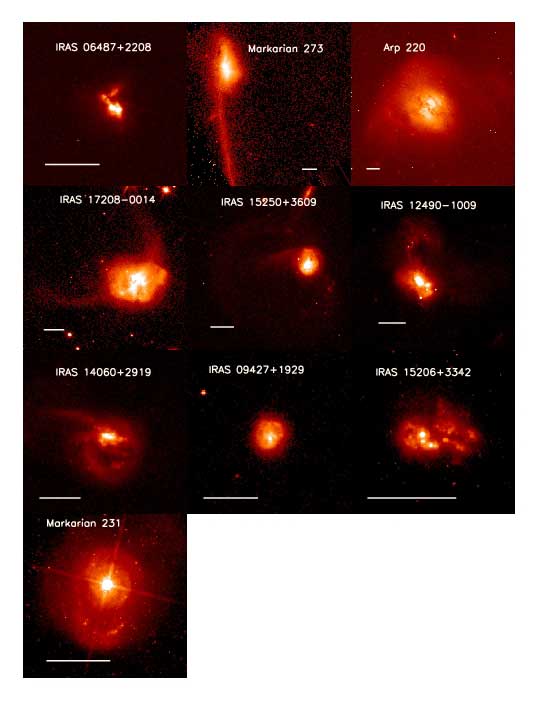}
\caption{\textbf{cont}. Post-coalescence ULIRGs.}\label{galaxias2WFPC2}
\end{figure*}

\
\begin{table*}[t]
\begin{minipage}[t]{\textwidth}
\centering
\renewcommand{\footnoterule}{}  
\caption{Main characteristics of the sample of ULIRGs. The galaxies are sorted by decreasing nuclear distance.}   
\label{TableSample}      
\tabcolsep0.1cm
\begin{tabular}{lcccccccc}        
\noalign{\smallskip}
\hline
\noalign{\smallskip}
\hline
\noalign{\smallskip}
IRAS Name &R.A.(J2000) & DEC(J2000) & Spectral Class & \textit{z} & D$_{\rm{L}}$\footnote{For the luminosity distances, the Wright (2006) cosmology calculator has been used.} & ${\log(L_{\rm{IR}})}$\footnote{L$_{IR} $(8-1000 $\mu$m) was derived following Sanders \& Mirabel 1996 ($L(8-1000\mu m)=4\pi {D_{L}}^{2}F_{IR}[L_{\odot}]$, with $F_{FIR}$=$1.26\times 10^{-14}\cdot(2.58f_{60}+f_{100})[Wm^{-2}]$. The quantities $f_{12}$, $f_{25}$, $f_{60}$, and $f_{100}$ are the \textit{IRAS} flux densities in Jy at 12, 25, 60 and 100 $\mu$m.). The IRAS fluxes were obtained from the IRAS Faint Galaxy Sample (Moshir et al.\,1993, Vizier on line catalogue II/156A).} & Scale & Morphology\footnote{IP means interacting pair, DN double nucleus and SN single nucleus. The distances in kpc represent the projected nuclear distance as measured in the F814W HST image.} \\
 & (hh:mm:ss) & (deg:mm:ss) & (Optical) & & (Mpc) &(L$_{\odot}$) &(kpc arcsec$^{-1}$) & \\
\hline
IRAS~13156+0435      &              13:18:07.6 & +04:18:59.4 & H\,{\sc ii}/LINER (G)& 0.113 & 525 & 12.13                & 2.058 & IP at 36.0~kpc \\
IRAS~13342+3932      &              13:36:23.3 & +39:16:41.9 & Sey1 (E)             & 0.179 & 866 & 12.51                & 3.026 & IP at 25.1~kpc  \\
IRAS~18580+6527      &              18:58:06.3 & +65:31:32.2 & H\,{\sc ii}/Sey2 (G) & 0.176 & 850 & 12.26   & 2.986 & IP at 15.0~kpc 	      \\
IRAS~16007+3743   &              16:02:36.9 & +37:34:39.7 & LINER (H)            & 0.185 & 898 & 12.11                & 3.100 & IP at 14.2~kpc	      \\
IRAS~06268+3509      &              06:30:12.6 & +35:07:51.9 & H\,{\sc ii} (C)\footnote{(C) Darling \& Giovanelli 2006}      & 0.169 & 813 & 12.51\footnote{IRAS fluxes obtained from the IRAS point Source Catalogue.}   & 2.895 & IP at 9.1~kpc \\
IRAS~08572+3915      &              09:00:25.4 & +39:03:54.4 & LINER (E)\footnote{(E) Kim, Veilleux \& Sanders 1998.}           & 0.058 & 259 & 12.17                & 1.130 & IP at 6.1~kpc \\
IRAS~14348-1447   &              14:37:38.4 & -15:00:22.8 & LINER (E)            & 0.083 & 378 & 12.39                & 1.556 & IP at 5.5~kpc 	      \\
Arp~299\footnote{Arp~299 is the system formed by NGC~3690 (to the east) and IC~694 (to the west).} & 11:28:30.4 & +58:34:10.0 & H\,{\sc ii}/Sey2 (D)\footnote{(D) Garc\'{\i}a-Mar\'{\i}n et al.\,2006.} & 0.010 & 43  & 11.81                & 0.205 & IP at 5.0~kpc  \\   
IRAS~13469+5833      &              13:48:40.3 & +58:18:50.0 & H\,{\sc ii}  (I)\footnote{Veilleux et al.\,1999.}     & 0.158 & 755 & 12.31                & 2.726 & IP at 4.5~kpc    \\
IRAS~12112+0305               &12:13:42.9 & +02:48:29.0 & LINER (E)            & 0.073 & 330 & 12.37                & 1.395 & IP at 4.0~kpc  \\
Mrk~463     & 13:56:02.8 & +18:22:17.2 & Sey1/Sey2 (F, G)\footnote{(F) Miller \& Goodrich 1990; (G) Garc\'{\i}a-Mar\'{\i}n 2007.}     & 0.050 & 222 & 11.81                & 0.984 & DN at 3.8~kpc    \\
IRAS~06487+2208      &              06:51:45.7 & +22:04:27.0 & H\,{\sc ii} (C)      & 0.144 & 682 & 12.57\footnote{IRAS fluxes obtained from the IRAS point Source Catalogue.} & 2.522 & DN at 1.5~kpc \\
IRAS~11087+5351      &              11:11:36.4 & +53:35:02.0 & Sey1 (G)             & 0.143 & 677 & 12.13                & 2.507 & DN at 1.5~kpc  \\
Mrk~273     & 13:44:42.0 & +55:53:12.1 & LINER/Sey2 (B, E)\footnote{(B) Colina, Arribas \& Borne 1999.}    & 0.038 & 167 & 12.18                & 0.749 & DN at 0.7~kpc    \\
Arp~220     & 15:34:57.1 & +23:30:11.5 & LINER (E)            & 0.018 & 78  & 12.20                & 0.368 & DN at 0.4~kpc	      \\ 
IRAS~17208-0014   &              17:23:21.9 & -00:17:00.9 & LINER (A)\footnote{(A) Arribas \& Colina 2003.}           & 0.043 & 190 & 12.43                & 0.844 & SN 	      \\
IRAS~15250+3609   &              15:26:59.4 & +35:58:37.5 & LINER (H)\footnote{(H) Veilleux et al.\,1995.}            & 0.055 & 245 & 12.09                & 1.072 & SN 	      \\
IRAS~12490-1009      &              12:51:40.7 & -10:25:26.1 & H\,{\sc ii}/LINER (G)& 0.101 & 465 & 12.07                & 1.854 & SN  \\
IRAS~14060+2919      &              14:08:18.9 & +29:04:46.9 & H\,{\sc ii}  (E)     & 0.117 & 545 & 12.18                & 2.113 & SN    \\
IRASF~09427+1929     &              09:45:32.4 & +19:15:34.9 & H\,{\sc ii} (H)      & 0.149 & 708 & 12.10                & 2.600 & SN \\
IRAS~15206+3342   &             15:22:38.0 & +33:31:35.9 & H\,{\sc ii} (E)      & 0.124 & 580 & 12.27                & 2.232 & SN 	      \\
Mrk~231     & 12:56:14.2 & +56:52:25.2 & Sey1 (E)             & 0.042 & 186 & 12.57                & 0.832 & SN  \\
\hline                                   
\end{tabular}          
\end{minipage}
\end{table*} 

\onltab{2}{
\begin{table}[t]
\begin{minipage}[t]{\columnwidth}
\centering
\renewcommand{\footnoterule}{}  
\caption{Characteristics of the INTEGRAL fiber bundles}   
\label{INTEGRAL-bundles}      
\tabcolsep0.1cm
\begin{tabular}{lcccc}        
\noalign{\smallskip}
\hline
\noalign{\smallskip}
\hline
\noalign{\smallskip}
Bundle & Fiber core ($\oslash$)\footnote{Fiber \textit{core} diameter (arcsec).} & \# Fibers\footnote{ Total number of fibers (rectangle+sky ring).} & FoV\footnote{Spatial coverage of the central rectangle in arcsec.} & Sky-ring\footnote{Diameter of the external sky ring in arcsec.}\\
\hline
SB1 & 0.45 & 205(175+30) & 7.80$\times$6.40 & 90\\
SB2 & 0.90 & 219(189+30) & 16.0$\times$12.3 & 90 \\
SB3 & 2.70 & 135(115+20) & 33.6$\times$29.4 & 90 \\
\hline                                   
\end{tabular}
\end{minipage}
\end{table}
}

   \begin{figure*}[t]
   \centering
   \includegraphics[width=\textwidth]{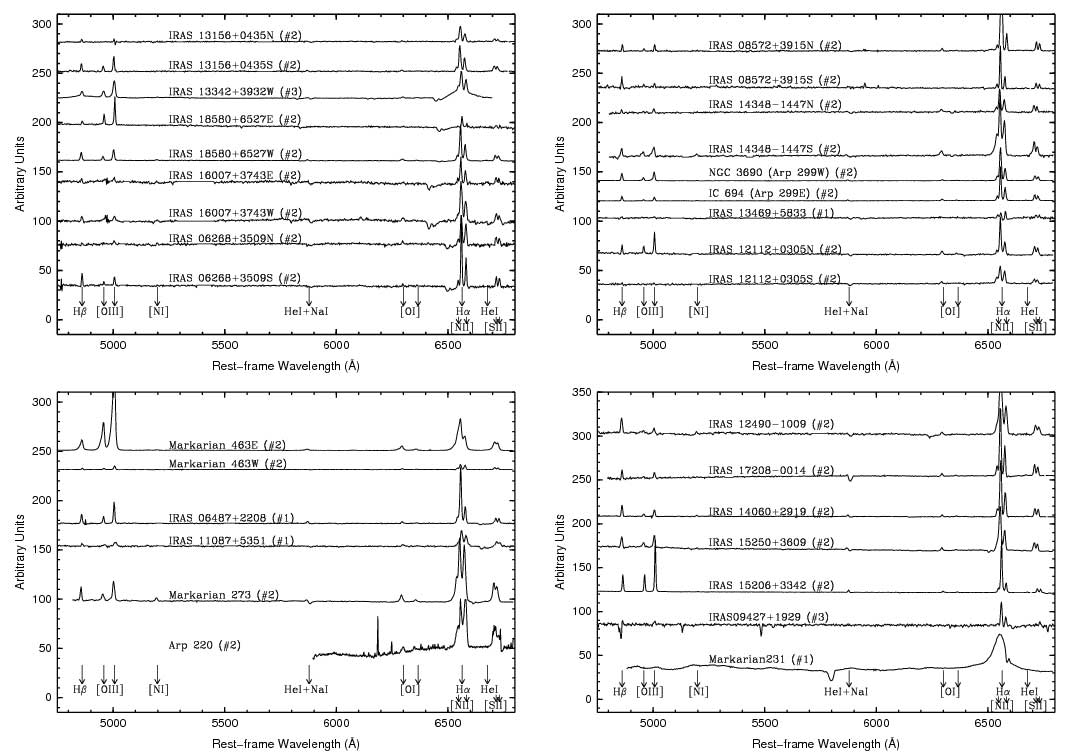}
   \caption{Nuclear spectra of the sample of galaxies. The number of the INTEGRAL standard bundle used is specified next to the galaxy name. Each spectrum corresponds to a region of 0\farcs45, 0\farcs9 and 2\farcs7, depending on the bundle used (SB1, SB2, and SB3 respectively). For Arp~220, we show the region of interest observed with the SB2 INTEGRAL bundle. For IRAS13343+3932, we only show the western nucleus spectrum, since the eastern one was of low S/N ratio. The integration time for IRAS13469+5833 was insufficient to achieve a good S/N ratio. Its H$\alpha$+[N\,{\sc ii}] complex was also strongly contaminated by sky lines. Hence, here we only show one spectrum as example.
}
         \label{espectros1}
   \end{figure*}

\section{Observations and data reduction}
IFS observations were obtained with INTEGRAL, a fiber-based optical integral field system (Arribas et al.\,1998) connected to the Wide Field Fibre Optic Spectrograph (WYFFOS; Bingham et al.\,1994) and mounted on the 4.2 m William Herschel Telescope. The observations were carried out during a number of observing runs between 1998 and 2004. We used three INTEGRAL configurations, namely the so-called standard bundles 1 (SB1), 2 (SB2), and 3 (SB3). The fiber diameters are 0\farcs45, 0\farcs9, and 2\farcs7 for SB1, SB2, and SB3, respectively, resulting in the field of view (FoV) given in Table\,\ref{INTEGRAL-bundles}. The three bundles have a similar configuration in the focal plane. The majority of the fibers form a rectangular area centered on the object, whereas a subset of fibers that form an outer ring of 45\arcsec\,in radius is simultaneously used for measuring the sky (see configuration details in Table \ref{INTEGRAL-bundles}). A remarkable capability of the instrument INTEGRAL is its flexibility in the bundle change, which allows us to select the most convenient instrument configuration depending on the seeing conditions. For the majority ($\sim$75\%) of our cases, the SB2 bundle was the preferred one: the fiber size (0\farcs9) is similar to the typical seeing of La Palma, and in general an entire ULIRG fits in the FoV (16\farcs0$\times$12\farcs3). 

The spectra were taken with a 600 lines mm$^{-1}$ grating, which provides an effective spectral resolution (FWHM) of 6.0, 6.0, and 9.8 \AA\,for the SB1, SB2, and SB3 bundles, respectively\footnote{These values correspond to the old camera mounted on WYFFOS, which was used for the present observations. In August 2004, a new camera was commissioned for the instrument. See more details in http://www.iac.es/proyecto/integral.}. The covered spectral range of interest was $\lambda\lambda$4500-7000 \AA\,(rest-frame). The total integration time, fiber bundle used on each galaxy, and individual comments on some observations are presented in Table \ref{datos}.

The data reduction was performed within the IRAF\footnote{IRAF software is distributed by the National Optical Astronomy Observatory (NOAO), which is operated by the Association of Universities for Research in Astronomy (AURA), Inc., in cooperation with the National Science Foundation.} environment, followed the standard procedures applied to this type of data (see Arribas et al.\,1997 and references therein), and can be summarized as follows. After subtracting an averaged BIAS frame, we proceeded to define the fibers apertures. This step involves the identification of information coming from each fiber, to preserve the full spatial and spectral information provided by IFS. For this reduction step, we used a sky flat image, which clearly indicates the position of spectra from the fibers along the detector because of its high signal over the entire wavelength range. The location of the INTEGRAL+WYFFOS system at the Nasmyth focus of the telescope ensures its stability throughout a given night (del Burgo 2000), thus the information obtained for the sky flat can be used in all the scientific images observed during the same night. Once the fibers have been identified, we proceed to subtract the stray light and cross-talk contamination. Both signals create a background that has a diluting effect on the spectra, affecting in particular spectra from low surface brightness regions. The spurious light outside the defined apertures was analyzed,  modeled two-dimensionally, and extracted from every image. The light transmitted by each fiber onto the detector was afterwards measured, added along the spatial direction, and subtracted to obtain a final one dimensional spectrum for every single fiber.

 The one dimensional spectra were individually wavelength calibrated using a well-characterized lamp arc image. We checked the wavelength calibration using well identified sky lines, and obtained standard deviations of between 0.08 and 0.18 \AA\,. 

The next step in the data reduction involves the flat-field correction. By combining the information provided by lamp and sky flats, which accounted for pixel-to-pixel and fiber-to-fiber variations, respectively, we derived a response image, which was applied to all science data. Finally, the sky contribution to the spectra was defined and subtracted using the combined information of the outer ring of fibers present in every bundle. We carefully checked that the fibers covering the sky were not contaminated by any contribution of the galaxy. 

For the flux calibration, several spectrophotometric standard stars\footnote{These stars are part of the Hubble Space Telescope spectrophotometric standards.} were observed (BD+28~4211, Feige~34, BD +3326~42, and GD~153) using the same instrumental configuration and data reduction procedures as for the galaxies. Once the data reduction was completed, we combined all the individual exposures (a minimum of three), improving the S/N and rejecting cosmic rays. Finally, when needed, a correction factor accounting for the percentage of flux not integrated by the central fiber (i.e., the fiber where the calibration star is centered) was used. Using the four spectrophotometric standard stars during six different observing runs, the average uncertainty obtained in the flux calibration of the galaxies was in the range 10--15\%. For further details about the flux calibration technique, we refer to Monreal-Ibero et al. (2007).

We also retrieved broad-band imaging data for our galaxies from the \textit{HST} archive. The images were taken with the Wide Field and Planetary Camera (WFPC2) using the F814W filter, and are available for the entire sample. This filter is equivalent to the ground-based Johnson-Cousins I (Origlia \& Leitherer 2000). The images were calibrated on the fly, with the highest quality available reference files at the time of retrieval\footnote{More information about data process can be found at www.stsci.edu/hst/wfpc2/analysis/analysis.html}.

\begin{table*}[!t]
\begin{minipage}[t]{\textwidth}
\centering
\renewcommand{\footnoterule}{}  
\caption{Integral field spectroscopy observations log. The galaxies are sorted by decreasing nuclear distance.}   
\label{datos}      
\tabcolsep0.1cm
\begin{tabular}{lccccc}        
\noalign{\smallskip}
\hline
\noalign{\smallskip}
\hline
\noalign{\smallskip}
IRAS Name & INTEGRAL & INTEGRAL & Date & Exp. time & Comment on IFU data\\
 & (Bundle) & (FoV arcsec$^{2}$) &month/year & (s) & \\
\hline
IRAS~13156+0435  & SB2, SB3 & 32.9$\times$25.3, 69.1$\times$60.5& 03/02 & 18000\footnote{3$\times$1500 s for the South component with the SB2 bundle; 3$\times$1800 for the North component with the SB2 bundle, 5$\times$1500 s for the entire system with the SB3 bundle.}    & [O\,{\sc iii}] non detected in N comp. \\
IRAS~13342+3932  & SB3   & 101.7$\times$89.0& 03/02 & 6$\times$1500     &                   \\
IRAS~18580+6527   & SB2   & 47.8$\times$36.7 & 04/01 & 7$\times$1500     &                    \\
IRAS~16007+3743   & SB2   & 49.6$\times$38.1 & 04/01 & 6$\times$1500     &                    \\
IRAS~06268+3509 & SB2  & 46.3$\times$35.6 & 03/02 & 6$\times$1500      &                   \\
IRAS~08572+3915\footnote{Previously published galaxy. See Colina, Arribas \& Monreal-Ibero 2005, and references therein.}  & SB2   & 18.0$\times$13.9 & 04/99 & 9$\times$1800      &                    \\
IRAS~14348$-$1447$^{a}$ & SB2   & 1.3$\times$1.0& 04/98 & 4$\times$1800     &                    \\ 
Arp~299 \footnote{Garc\'{\i}a-Mar\'{\i}n et al. 2006.}  & SB2, SB3 & 3.3$\times$2.5, 6.9$\times$6.0& 01/04 & 14400\footnote{3$\times$1200 s for each individual galaxy of the system, NGC~3690 and IC~694 observed with the SB2 bundle. 3$\times$1800 s for the entire system as observed with the SB3 bundle.}  & \\
IRAS~13469+5833  & SB1   & 21.3$\times$17.4& 03/02 & 5$\times$1800     & Sky line to H$\alpha$+[N\,{\sc ii}] complex\\
         &     & &    &            & Not enough S/N                        \\
IRAS~12112+0305$^{a}$\,\footnote{For IRAS~12112+0305 we present the results obtained by combining two different pointings of the SB2 INTEGRAL bundle.}  & SB2   & 27.9$\times$19.5& 04/98 & 5$\times$1800, 4$\times$1500    & \\
Mrk~463      & SB2   & 15.7$\times$12.1& 04/01 & 5$\times$900      &                    \\
IRAS~06487+2208  & SB1   & 19.7$\times$16.1 & 03/02 & 5$\times$1500      &                    \\
IRAS~11087+5351  & SB1   & 19.5$\times$16.0& 03/02 & 6$\times$1500     &                    \\
Mrk~273$^{a}$      & SB2   & 12.0$\times$9.2& 04/98 & 3$\times$1500     &                    \\
Arp~220$^{a}$ & SB2, SB3 & 5.9$\times$4.5, 12.4$\times$10.8 & 05/00 & 48000\footnote{6$\times$1500 s for the nuclear region observed with the SB2 bundle, 26$\times$1500 s for the nucleus and the extended nebula as observed with the SB3 bundle.} & [O\,{\sc iii}] not detected, H$\beta$ not covered \\	
IRAS~17208$-$0014$^{a}$ & SB2   & 13.5$\times$10.4 & 04/99 & 4$\times$1950     &                    \\
IRAS~15250+3609$^{a}$  & SB2   & 17.1$\times$13.2& 04/98 & 5$\times$1800     &                    \\
IRAS~12490$-$1009 & SB2   & 29.7$\times$22.8& 03/02 & 4$\times$1500     &                    \\
IRAS~14060+2919  & SB2   & 1.9$\times$1.4& 03/02 & 4$\times$1800     &                    \\ 
IRAS~09427+1929  & SB3   & 87.4$\times$76.4 & 03/02 & 2320\footnote{1500 s, 520 s, 300 s.} & Short inhomogeneous exposures\\
IRAS~15206+3342$^{a}$   & SB2   & 35.7$\times$27.4& 04/99 & 4$\times$1800     &                    \\ 
Mrk~231      & SB1   & 6.5$\times$5.3& 03/02 & 4$\times$900      & H$\beta$ not covered         \\ 
         &     & &    &            & Broad emission lines          \\
\hline                                   
\end{tabular}\label{Table_log}
\end{minipage}
\end{table*}

\section{Data analysis}
After data reduction, every galaxy has a set of spectra, each of them associated with a particular galaxy region observed by an individual fiber. The nuclear spectra of the systems under study are shown in Fig.\,\ref{espectros1}. Depending on the instrument set-up, the angular scale of the spectra is 0\farcs45, 0\farcs9, or 2\farcs7 for the SB1, SB2, and SB3 bundles, respectively. In this context, nuclear spectra are the ones corresponding to the fiber closest to the continuum peak. 

Almost all galaxies exhibit nuclear emission, the most important optical emission lines including those that are both strong (H$\beta$, [O\,{\sc iii}]$\lambda\lambda$4959, 5007, H$\alpha$+[N\,{\sc ii}]$\lambda\lambda$6548, 6484, and [S\,{\sc ii}]$\lambda\lambda$6716, 6731) and weak ([O\,{\sc i}]$\lambda$6300). Some galaxies also exhibit other weak emission lines ([N\,{\sc i}]$\lambda$5199, He\,{\sc i}$\lambda$5876, and He\,{\sc i}$\lambda$6678), as well as interstellar absorption lines (Na\,{\sc i}$\lambda\lambda$5890, 5896).

To obtain maps of the most relevant optical lines tracing the ionization state, we first fitted the lines to Gaussian functions using the \texttt{DIPSO} package (Howarth \& Murray 1988), inside the STARLINK environment\footnote{See http://www.starlink.rl.ac.uk/.}. It is well known that, under certain conditions, such as line blending, the fitting algorithms may derive different solutions. We therefore decided to apply some restrictions to constrain our results more reliably. The H$\beta$ and [O\,{\sc i}]$\lambda$6300 lines were individually fitted with a single component, with no particular constraint. We note that no H$\beta$ in absorption was detected, and hence no correction was applied. The [O\,{\sc iii}]$\lambda\lambda$4959, 5007, and H$\alpha$+[N\,{\sc ii}] complexes were fitted with two and three Gaussians, respectively, which were assumed to have identical kinematics. We applied these conditions based on the assumption that these line complexes trace identically the ionized gas kinematics. Additionally, we fixed the line intensity ratios of the [O\,{\sc iii}] and [N\,{\sc ii}] lines according to atomic parameters. For the [S\,{\sc ii}]$\lambda\lambda$6716, 6731 lines, we assume only that they share the same kinematics, since their line ratio variations can be used to trace the electron density.

The spectral maps shown in this paper have all been derived using the values obtained with a single Gaussian component fitting for each line. In few cases, we detected additional kinematical components in the line profiles, which may be indicative of the presence of winds, superwinds, and AGN flows. Some regions of a few galaxies are also consistent with the presence of a broad component in the hydrogen recombination lines. These particular cases will not be addressed here but in dedicated papers about the kinematics and excitation conditions, because the purpose of the present work is to provide a general overview of the data.

The mappings tracing the stellar component were derived using a continuum emission line-free filter with a rectangular bandwidth of about 150 \AA\, for each galaxy (filters being centered approximately at 4700 \AA\, and 6150 \AA\, restframe).

Finally, the derivation of all these maps involves the ordering of the fibers according to their astronomical position (RA, Dec). To that aim, we used the free software IDA (Garc\'{\i}a-Lorenzo et al.\,2002), a tool specifically designed  for the INTEGRAL IFU. To preserve the instrumental spatial resolution, the pixel size was equivalent to the fiber size used during the observations (see Figs.\,\ref{maps1} - \ref{maps28}). The HST images and the spectral maps were aligned by comparing with the peaks and external envelope structure of the red continuum map.    
 
\section{Observed morphologies of the stellar and ionized gas components}
\subsection{General trends}
 We present and discuss the differences between the structure of the stellar and ionized gas components for our sample of ULIRGs. 

The high resolution HST images provide evidence for very complex stellar structures, with bright nuclear regions, dust structures, and knots of star formation. In the pre-coalescence systems, the large scale structure is characterized by merger-induced structures, i.e., bridges, tails, and plumes connecting the individual galaxies. The original spiral arms are disrupted and form kpc-size tidal tails with, in some cases, chains of star-forming regions (Monreal-Ibero et al.\,2007). For the post-coalescence galaxies, the large scale structure traces an outer common envelope, which resembles an elliptical structure (Garc\'{\i}a-Mar\'{\i}n\,2007), in addition to late merger features such as disordered nuclear regions and tails. 

In spite of the different angular sampling, the INTEGRAL-based continuum images of the galaxies obtained at wavelengths blueward of H$\alpha$ and H$\beta$ recover the stellar structure observed in the high angular resolution \textit{HST} images. This occurs even in the low-surface brightness regions, reassuring us that all the observed stellar and ionized gas structures are real. Only small differences between the INTEGRAL red and blue continua are detected (e.g., Arp~299, IRAS~08572+3915, IRAS~12112+0305), and they can probably be explained by extinction effects.  

 The general morphology of the ionized gas low surface brightness regions is similar to that of the stellar component, although in about 25\% of the galaxies structural variations such us differences in shape or the presence of external clumps are associated with gas in high excitation states (Arp~299, Mrk~463, Mrk~273, IRAS~18580+6527, and IRAS~11087+5351). These differences are mainly traced by the [O\,{\sc iii}]$\lambda$5007. In the pre-coalescence systems, one of the optical nuclei is usually a weak line emitter, with a surface brightness similar to that of the interface between the galaxies.

The largest structural variations between the stellar and ionized gas components are observed in the high surface brightness regions (i.e., the main body of the galaxies), and are detected as differences in the location of the emission peaks (see the distance between continuum and emission lines peaks for individual ULIRGs in Table\,\ref{Separacion}). In what follows, we consider a significant shift as those equal to or larger than 1~kpc (within the errors given in Table\,\ref{Separacion}). 

There is also evidence of substructures in the different ionization states of the gas. These variations may be due to mechanisms such as star formation, AGNs, and shocks caused by tidal effects. To trace this more reliably, we have selected respectively the strong hydrogen recombination line H$\alpha$, the [O\,{\sc iii}]$\lambda$5007 line, which traces hard ionizing photons, and [O\,{\sc i}]$\lambda$6300, which is considered to be a sensitive shock tracer. Other effects, such as extinction and differences in stellar populations and in metallicities, may also play a role, but the detailed analysis of the excitation mechanisms in ULIRGs is beyond the scope of the present paper. In the following, we present the main structural characteristics of the pre- and post-coalescence systems. 

Some ULIRGs of the sample were not included for a variety of reasons, such as the lack of [O\,{\sc iii}]$\lambda$5007 detection (Arp~220), the low S/N of their data precluding the detection of emission lines (e.g., IRAS~13469+5833), poor spatial resolution (e.g., IRAS~13342+3932), and severe AGN contamination of the host galaxy light that makes it impossible to distinguish the ionized gas component (Mrk~231, see Appendix). The analysis presented here was carried out for 18 ULIRGs (i.e., 20 INTEGRAL pointings, see Table\,\ref{Table_log}). This includes two individual pointings for the spatially separated galaxies that form the systems Arp~299 and IRAS~13156+0435.
On average, the linear scale corresponding to our angular resolution is about 1.8 kpc. Therefore, all the results presented here should be considered valid on this scale. 

\subsection{Pre-coalescence ULIRGs}
The ionized gas traced by the H$\alpha$ maps in the pre-coalescence systems exhibit emission on scales of about 5-10 kpc measured from the stellar nuclei (see Figs, \ref{maps1} - \ref{maps16}). Their intensity peak are located at different positions: in the brightest stellar nuclei (e.g., IRAS~18580+6527), in the main body/overlapping region between the galaxies (e.g., NGC~3690), in one of the tails (e.g., IRAS~16007+3743), in an extra-nuclear stellar knot (e.g., IRAS~12112+0305), or in the secondary nucleus (e.g.\,Mrk~463). In about 65\% of the galaxies studied, there are significant offsets between the H$\alpha$ and the stellar continuum peaks, with projected separations of between 1-8~kpc. Extra-nuclear star-forming regions, mainly located in tidal tails, are also detected as secondary H$\alpha$ peaks (IRAS~16007+3743, IRAS~08572+3915, IRAS~14348-1447, and IRAS~12112+0305). These additional light peaks are identified as Tidal Dwarf Galaxies (TDGs), candidates (Monreal-Ibero et al.\,2007). The H$\beta$ structure is rather similar to that of H$\alpha$, but it is intrinsically weaker and more affected by extinction, especially in the galaxy nuclei. 

 The overall morphology, relative brightness of the nuclei, and the peak distribution of the brightest forbidden lines [N\,{\sc ii}]$\lambda$6484, and [S\,{\sc ii}]$\lambda\lambda$6716, 6731 are in good agreement with that of H$\alpha$. Nonetheless, there are some cases where small scale differences in their peak positions are found (IRAS~18580+6527 and IRAS~12112+0305), and one galaxy where their peaks are located 8.4~kpc away from the one of H$\alpha$ (IRAS~16007+3743). Given the wavelength proximity of these emission lines, it is unlikely that these differences are produced by any extinction effects. They are probably produced instead by the presence of different ionization mechanisms associated with tidal shocks. 

The results for the highly excited line [O\,{\sc iii}]$\lambda$5007 are an interesting case study. As for H$\alpha$, in about 50\% of the pre-coalescence systems, there are significant shifts between the stellar continuum and the nebular emission peaks. In comparison with other emission-line maps, we find that the [O\,{\sc iii}]$\lambda$5007 peak coincides with that of H$\alpha$ in 70\% of the studied cases. In some galaxies, there are regions that appear to be dominated by local ionization sources. For instance, in IRAS~08572+3915 and IRAS~12112+0305, the [O\,{\sc iii}]$\lambda$5007 line has its maxima in regions identified as TDG candidates in the stellar continuum images. These TDG candidates are also secondary H$\alpha$ emitters. These differences may be caused by the effects of extinction on [O\,{\sc iii}]$\lambda$5007, which are more important in the nuclei than in these external regions. In the case of IC~694, the peak of the [O\,{\sc iii}]$\lambda$5007 coincides with a region of the galaxy believed to be one of the original spiral arms disrupted by the merger. It is therefore also bright in the stellar component and in H$\alpha$. 

The galaxy IRAS~18580+6527 also deserves special mention, since the extra-nuclear local maxima located at about 10~kpc from the eastern nucleus (also bright in H$\alpha$, see Fig.\,\ref{maps5}, $\Delta$$\alpha$$\sim$-4, $\Delta$$\delta$$\sim$-3) is not associated with any particular stellar mass concentration, but is rather dominated by nebular emission. Interestingly enough, this galaxy is classified as a Seyfert. As explained below, similar extra-nuclear highly excited nebulae are detected in the post-coalescence ULIRGs Mrk~273 and IRAS~11087+5351.

Approximately 40\% of the pre-coalescence ULIRGs present shifts of up to 6~kpc between the peaks of the red continuum and the [O\,{\sc i}]$\lambda$6300 line. However, in some cases such as IRAS~06268+3509 and Mrk~463, the line maxima is associated with the stellar secondary nucleus of the system. This would mean that in general the [O\,{\sc i}]$\lambda$6300 activity is mainly related to the nuclei.

\subsection{Post-coalescence ULIRGs}

For the morphologically evolved ULIRGs, the comparison between the stellar and ionized gas components provides different results from above. The most important one is that all the emission tends to be concentrated in the nuclear region. The previously reported structural differences between the stellar and ionized gas are not so common for this type of ULIRGs. The spectral maps indicate that in all cases (see Figs. \ref{maps17} - \ref{maps28}) the H$\alpha$ peak coincides with the peak of the stellar emission (i.e. within the central kpc), indicating a nuclear concentration of the excitation sources. The H$\alpha$ overall structure generally coincides with that of the red continuum. One galaxy (IRAS~15250+3609) contains a TDG candidate that is also an H$\alpha$ emitter (Monreal-Ibero 2007). For Mrk~273, Arp~220, and IRAS~11087+5351, there are also secondary extra-nuclear structures that do not correlate spatially with the stellar distribution. Likewise for the pre-coalescence systems, the H$\alpha$ map generally agree with those of H$\beta$, although the latter is more affected by extinction that can cause their maxima position not to be coincident.

The behavior of the [N\,{\sc ii}]$\lambda$6484, and [S\,{\sc ii}]$\lambda\lambda$6716, 6731 follows that of H$\alpha$, and only small scale variations are measured in IRAS~15250+3609.

As in the pre-coalescence systems, there are two galaxies (IRAS~11087+5351 and Mrk~273) that exhibit a peak in the [O\,{\sc iii}]$\lambda$5007 emission within the extra-nuclear regions (7.2 and 5.1~kpc, respectively). These regions are not dominated by local ionization sources, because there is no particular stellar mass concentration there. Interestingly, both galaxies have a Seyfert classification (see Table \ref{TableSample}). Secondary H$\alpha$ structures are also detected in these extra-nuclear regions.

The [O\,{\sc i}]$\lambda$6300 emission line maps also follow the nuclear concentration trend exhibited by the post-coalescence systems, with only the maxima position of IRAS~11087+5351 being shifted with respect to that of the continuum and almost coinciding with the [O\,{\sc iii}]$\lambda$5007 peak.

\begin{table*}[!h]
\begin{minipage}[t]{0.8\textwidth}
\centering
\renewcommand{\footnoterule}{}  
\caption{Separation between the red continuum and the emission line gas peaks.}   
\label{Separacion}      
\tabcolsep0.1cm
\begin{tabular}{lcccccccc}        
\noalign{\smallskip}
\hline
\noalign{\smallskip}
\hline
\noalign{\smallskip}
Galaxy            & Nuc Sep &[O\sc{i}]    & H$\alpha$    & [O\sc{iii}]  & H$\beta$     & [S\sc{ii}]   & [N\sc{ii}] & Comment \\
                  & (kpc)   & (kpc)       & (kpc)        & (kpc)        & (kpc)        & (kpc)        & (kpc)      &         \\
\hline
IRAS~13156+0435N  & 36.0    & 1.8$\pm$0.9 & 1.8$\pm$0.9  & 2.0$\pm$0.9  & 1.8$\pm$0.9  & 1.8$\pm$0.9* & 1.8$\pm$0.9 & \tiny{Activity concentrated in the nucleus}\\                                         
IRAS~13156+0435S  & 36.0    & 1.8$\pm$0.9 & 1.8$\pm$0.9  & 1.8$\pm$0.9  & 1.8$\pm$0.9  & 1.8$\pm$0.9  & 1.8$\pm$0.9 & \tiny{Activity concentrated in the nucleus}\\                                         
IRAS~18580+6527   & 15.0    & 0.0$\pm$1.2 & 2.7$\pm$1.2  & 2.7$\pm$1.2  & 2.7$\pm$1.2  & 0.0$\pm$1.2  & 2.7$\pm$1.2 & \tiny{Secondary [O\,{\sc iii}] structure.}\\                                          
IRAS~16007+3743   & 14.2    & 3.9$\pm$1.4 & 8.4$\pm$1.4  & 8.4$\pm$1.4  & 8.4$\pm$1.4  & 0.0$\pm$1.4  & 0.0$\pm$1.4 & \tiny{H$\alpha$, [O\,{\sc i}] and [O\,{\sc iii}] offsets. TDG candidates.}\\          
IRAS~06268+3509   & 9.1     & 5.8$\pm$1.3 & 2.6$\pm$1.3  & 2.6$\pm$1.3  & 3.7$\pm$1.3  & 2.6$\pm$1.3  & 2.6$\pm$1.3 & \tiny{[O\,{\sc i}] peaks on secondary nucleus.}\\                                     
IRAS~08572+3915   & 6.1     & 5.9$\pm$0.5 & 5.9$\pm$0.5  & 5.1$\pm$0.5  & 1.0$\pm$0.5  & 5.9$\pm$0.5  & 5.9$\pm$0.5 & \tiny{H$\alpha$ peaks in secondary nucleus.}\\
  &    & &   &  &   & & & \tiny{ Extended [O\,{\sc iii}] nebula. TDG candidates.}\\
IRAS~14348$-$1447 & 5.5     & 0.0$\pm$0.7 & 0.0$\pm$0.7  & 0.0$\pm$0.7  & 0.0$\pm$0.7  & 0.0$\pm$0.7  & 0.0$\pm$0.7 & \tiny{TDG candidates.}\\                                                              
Arp~299/NGC~3690  & 5.0     & 0.4$\pm$0.1 & 1.3$\pm$0.1  & 1.2$\pm$0.1  & 1.2$\pm$0.1  & 1.2$\pm$0.1  & 1.3$\pm$0.1 & \tiny{H$\alpha$, [O\,{\sc i}] and [O\,{\sc iii}] offset.}\\                           
Arp~299/IC~694    & 5.0     & 0.8$\pm$0.1 & 0.9$\pm$0.1  & 1.3$\pm$0.1  & 1.3$\pm$0.1  & 0.9$\pm$0.1  & 0.9$\pm$0.1 & \tiny{H$\alpha$, [O\,{\sc i}] and [O\,{\sc iii}] offset.}\\                           
IRAS~12112+0305   & 4.0     & 1.3$\pm$0.6 & 2.0$\pm$0.6  & 1.2$\pm$0.6  & 2.5$\pm$0.6  & 1.2$\pm$0.6  & 1.2$\pm$0.6 & \tiny{H$\alpha$ and [O\,{\sc iii}] offsets. TDG candidates.}\\                                                                                             
Mrk~463           & 3.8     & 3.5$\pm$0.4 & 3.5$\pm$0.4  & 3.5$\pm$0.4  & 3.5$\pm$0.4  & 3.5$\pm$0.4  & 3.5$\pm$0.4 & \tiny{Ionized gas peaks in secondary nucleus.}\\                        
IRAS~06487+2208   &1.5      & 1.1$\pm$0.6 & 1.1$\pm$0.6  &  1.1$\pm$0.6 & 0.0$\pm$0.6  & 0.0$\pm$0.6  & 1.1$\pm$0.6 & \tiny{Activity concentrated in the nucleus}\\                                      
IRAS~11087+5351   &1.5      & 6.6$\pm$0.6 & 1.6$\pm$0.6  &  7.2$\pm$0.6 & 7.2$\pm$0.6  & 1.6$\pm$0.6  & 1.6$\pm$0.6 & \tiny{Extended [O\,{\sc iii}] nebula.}\\                                          
Mrk~273           &0.7      & 0.7$\pm$0.3 & 0.7$\pm$0.3  &  5.1$\pm$0.3 & 5.1$\pm$0.3  & 0.7$\pm$0.3  & 0.7$\pm$0.3 & \tiny{Extended [O\,{\sc iii}] nebula.}\\                                              
Arp220            &0.4      & 0.0$\pm$0.2 & 0.5$\pm$0.2  &  N/A         &    N/A       & 0.3$\pm$0.2  & 0.3$\pm$0.2 & \\                                              
IRAS~12490$-$1009 &0.0      & 1.7$\pm$0.8 & 1.7$\pm$0.8  &  0.0$\pm$0.8 & 0.0$\pm$0.8  & 1.7$\pm$0.8  & 1.7$\pm$0.8 & \tiny{Activity concentrated in the nucleus}\\                                         
IRAS~14060+2919   &0.0      & 1.9$\pm$0.9 & 1.9$\pm$0.9  &  1.9$\pm$0.9 & 1.9$\pm$0.9  & 1.9$\pm$0.9  & 1.9$\pm$0.9 & \tiny{Activity concentrated in the nucleus}\\                                         
IRAS~15206+3342   &0.0      & 0.0$\pm$1.0 & 0.0$\pm$1.0  &  0.0$\pm$1.0 & 0.0$\pm$1.0  & 0.0$\pm$1.0  & 0.0$\pm$1.0 & \tiny{Activity concentrated in the nucleus}\\                                         
IRAS~15250+3609   &0.0      & 0.0$\pm$0.5 & 1.4$\pm$0.5  &  1.4$\pm$0.5 & 1.4$\pm$0.5  & 0.0$\pm$0.5  & 0.0$\pm$0.5 & \tiny{Activity concentrated in the nucleus.\tiny{TDG candidate.}}\\           
IRAS~17208$-$0014 &0.0      & 0.0$\pm$0.4 & 0.0$\pm$0.4  &  1.1$\pm$0.4 & 1.1$\pm$0.4  & 0.0$\pm$0.4  & 0.0$\pm$0.4 & \tiny{Activity concentrated in the nucleus.}\\                                        
\hline                                  
\end{tabular}                                                                                                       
\end{minipage}
\end{table*}

\section{Summary}
This is the first paper in a series to study in detail the two-dimensional morphology, dust distribution, excitation processes, and kinematics of a representative sample of 22 local ULIRGs. To achieve this, we have used optical IFS data combined with high resolution \textit{HST} images. In this paper, we have described the sample selection and the observations, and presented the morphological properties of the continuum and ionized gas components. The ULIRGs were selected to cover a wider range of the IR luminosity distribution, different merger phases, and to exhibit a variety of nuclear activities (H\,{\sc ii}, LINER, Seyfert). Using the projected nuclear distance as a criteria, we have adopted a simple classification scheme that divides the sample into pre- and post-coalescence galaxies, and identifies the different behaviors at different stages of the merger. The main results of this paper can be summarized as:

\begin{itemize}
\item Despite the different resolution, the structure of the INTEGRAL red continua is consistent with that of the HST F814W band. The red and blue continua may exhibit slight differences caused by the effect of the extinction. In contrast, the structure of the warm ionized gas can be significantly decoupled from that of the stellar continuum. These variations are due to the different spatial distribution of the ionization sources present in the galaxy structure, and to the dust distribution. 

\item We have compared the stellar and ionized gas structures of ULIRGs undergoing different merger phases (pre- and post-coalescence), and found structural variations between the gaseous and the stellar components, with offsets in the emission peak positions of up to about 8~kpc in the pre-coalescence systems. For 64\% of ULIRGs in an early interaction phase, the H$\alpha$ peak does not coincide with the stellar maxima. In contrast, galaxies undergoing advanced mergers have their H$\alpha$ peak located in the nuclear regions. 

\item Four pre- and one post-coalescence ULIRGs appear to contain TDGs candidates, which we identify with H$\alpha$ emission associated with stellar emission knots.

\item The ionization structure traced by different emission lines has been also studied. We detected variations in lines with similar rest-frame, meaning that the differences are unlikely to be explained by the presence of dust. This is more common in the pre-coalescence ULIRGs, and it is probably a consequence of different excitation mechanisms. Differences in metallicity and stellar populations may also be playing a role.

\item The [O\,{\sc iii}] line traces highly excited extra-nuclear clouds in both pre- and post-coalescence systems, with no relevant stellar counterpart. These regions are also secondary H$\alpha$ emitters. All these galaxies have a Seyfert nuclei, classified using optical emission lines ratios.

\item The peak of the [O\,{\sc i}] line is shifted with respect to the stellar one, but in general its emission is nuclear. 

\item This analysis infers that there is an evolutionary trend in the ionized gas behavior. In the pre-coalescence ULIRGs, the structural differences are remarkable, and the extra-nuclear regions, out to distances of about 8~kpc, play an important role and contribute to the excitation level. In contrast, the post-coalescence ULIRGs tend to have their activity concentrated in the nucleus. The only exception to this are the extra-nuclear nebulae traced by the [O\,{\sc iii}]. We have detected them in both galaxy types, and they appear to be related to the presence of a Seyfert nucleus. 

\end{itemize}


\begin{acknowledgements}
We would like to thank Almudena Alonso-Herrero for useful comments and discussions. This paper uses the plotting package \texttt{jmaplot}, developed by Jes\'us Ma\'{\i}z-Apell\'aniz. http:$//$dae45.\break iaa.csic.es:8080$/$$\sim$jmaiz/software. This research has made use of the NASA/IPAC Extragalactic Database (NED) which is operated by the Jet Propulsion Laboratory, California Institute of Technology, under contract with the National Aeronautics and Space Administration.

This work has been supported by the Spanish Ministry of
Education and Science, under grant BES-2003-0852, project AYA2002-01055. MG-M is supported by the German federal department for education and research (BMBF) under the project numbers: 50OS0502 \& 50OS0801. AMI is supported by the Spanish Ministry of Science and Innovation (MICINN) under the program "Specialization in International Organisms" ref. ES2006-0003.
\end{acknowledgements}

\Online

\begin{appendix} 
\section{Stellar component and ionized gas maps for the sample of ULIRGs:}
In this Appendix, the maps of interest for all the galaxies of the sample, as obtained with the INTEGRAL system, are presented. The emission-line free stellar continuum, along with the most relevant optical-emission line (H$\beta$, [O\,{\sc iii}]$\lambda$5007, [O\,{\sc i}]$\lambda$6300, H$\alpha$+[N\,{\sc ii}]$\lambda\lambda$6548, 6584, [S\,{\sc ii}]$\lambda\lambda$)6716, 6731) maps are shown. Complementary optical (0.8 $\mu$m) \textit{HST} images are presented too. The galaxies are sorted by decreasing nuclear separation.

\begin{figure*}[t]
\centering
\includegraphics[width=\textwidth]{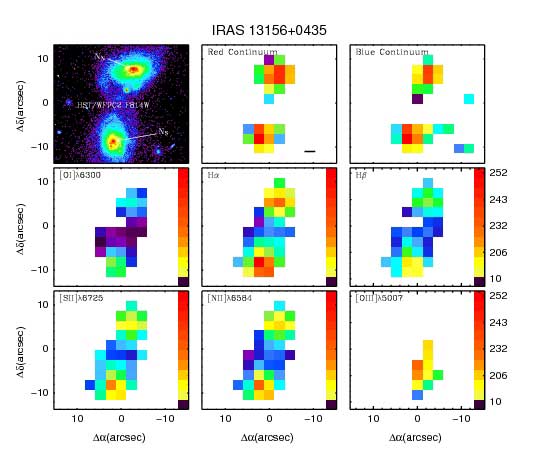} 
\caption{Emission-line free stellar continua and emission-line images of IRAS3156+0435 obtained with the INTEGRAL SB3 bundle. Contours represent the red continuum. The \textit{HST}-WFPC2 F814W image is shown for comparison. All the images are shown in a logarithmic scale. For a better comparison, the emission line maps have the same logarithmic color scale. The color code is given in relative flux units. North is up, east to the left. The scale represents 5 kpc. The arrows indicate the nuclei.}
\label{maps1}
\end{figure*}
\clearpage
\newpage

\begin{figure*}[h]
\centering
\includegraphics[width=\textwidth]{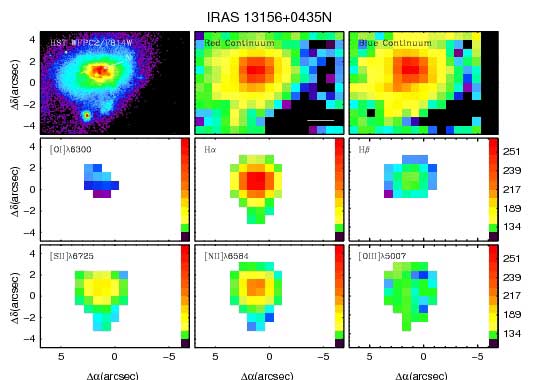} 
\caption{As Fig\,\ref{maps1} but for a galaxy observed with the SB2 bundle.}
\label{maps2}
\end{figure*}
\clearpage
\newpage

\begin{figure*}[h]
\centering
\includegraphics[width=\textwidth]{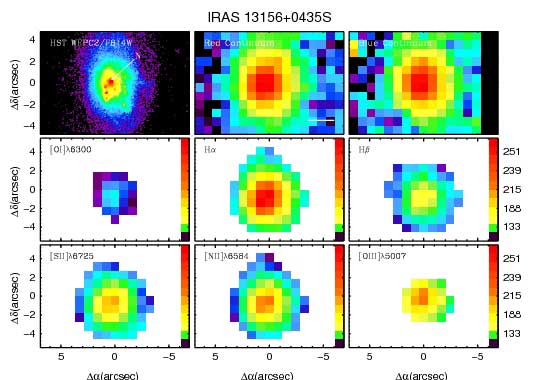} 
\caption{As Fig\,\ref{maps1} but for a galaxy observed with the SB2 bundle.}
\label{maps3}
\end{figure*}
\clearpage
\newpage

\begin{figure*}[h]
\centering
\includegraphics[width=\textwidth]{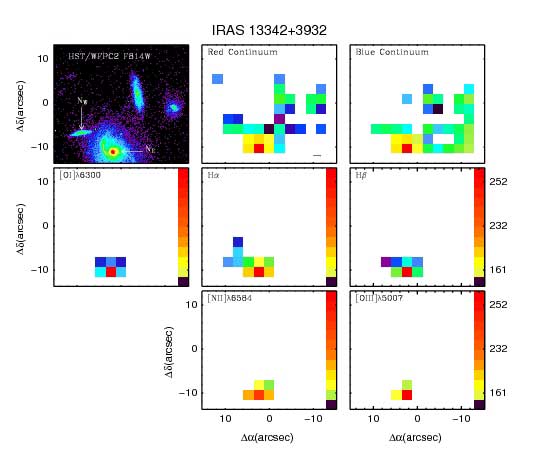} 
\caption{As Fig\,\ref{maps1}.}
\label{maps4}
\end{figure*}
\clearpage
\newpage

\begin{figure*}[h]
\centering
\includegraphics[width=\textwidth]{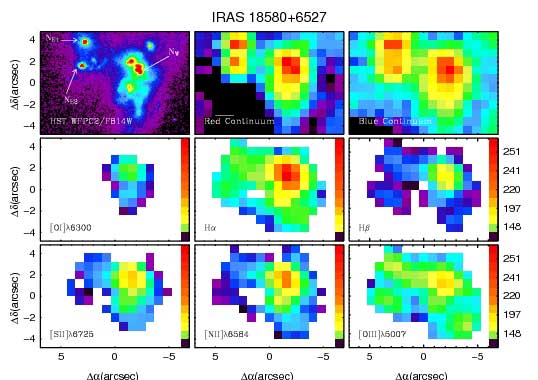} 
\caption{As Fig\,\ref{maps1} but for a galaxy observed with the SB2 bundle.}
\label{maps5}
\end{figure*}
\clearpage
\newpage

\begin{figure*}[h]
\centering
\includegraphics[width=\textwidth]{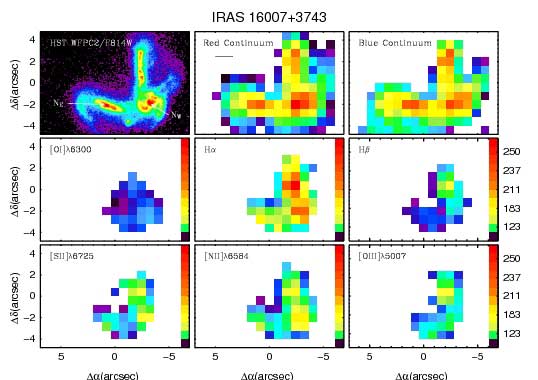} 
\caption{As Fig\,\ref{maps1} but for a galaxy observed with the SB2 bundle.}
\label{maps6}
\end{figure*}
\clearpage
\newpage

\begin{figure*}[h]
\centering
\includegraphics[width=\textwidth]{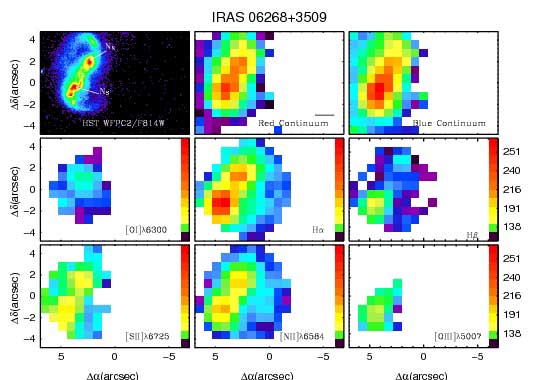} 
\caption{As Fig\,\ref{maps1} but for a galaxy observed with the SB2 bundle.}
\label{maps7}
\end{figure*}
\clearpage
\newpage

\begin{figure*}[h]
\centering
\includegraphics[width=\textwidth]{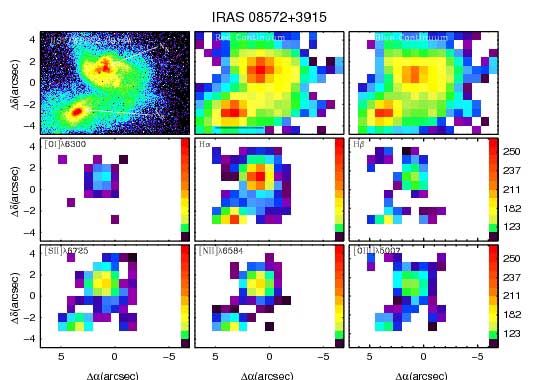} 
\caption{As Fig\,\ref{maps1} but for a galaxy observed with the SB2 bundle.}
\label{maps8}
\end{figure*}
\clearpage
\newpage

\begin{figure*}[h]
\centering
\includegraphics[width=\textwidth]{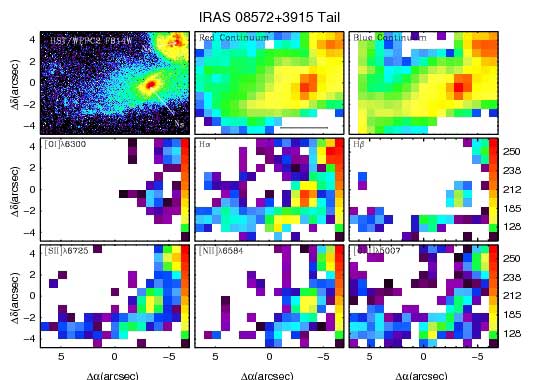} 
\caption{As Fig\,\ref{maps1} but for a galaxy observed with the SB2 bundle. This pointing covers the southern tail of IRAS~08572+3915.}
\label{maps9}
\end{figure*}
\clearpage
\newpage

\begin{figure*}[h]
\centering
\includegraphics[width=\textwidth]{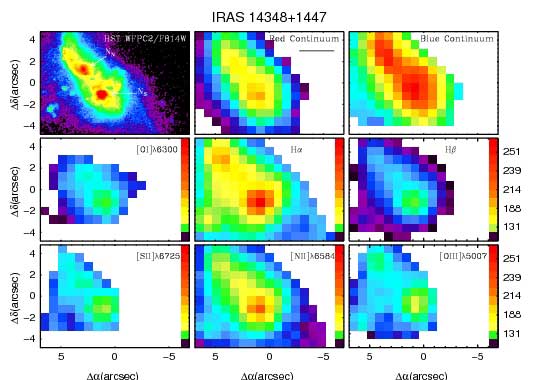} 
\caption{As Fig\,\ref{maps1} but for a galaxy observed with the SB2 bundle.}
\label{maps10}
\end{figure*}
\clearpage
\newpage

\begin{figure*}[h]
\centering
\includegraphics[width=\textwidth]{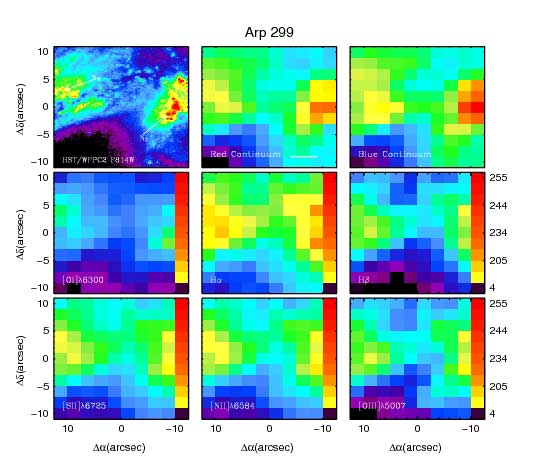} 
\caption{As Fig\,\ref{maps1}. In the Arp~299 system, NGC~3690 is located to the west, and IC~694 to the east. In this case the scale represents 2~kpc.}
\label{maps11}
\end{figure*}
\clearpage
\newpage

\begin{figure*}[h]
\centering
\includegraphics[width=\textwidth]{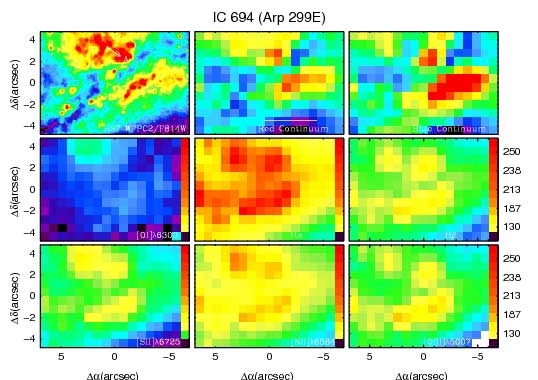} 
\caption{As Fig\,\ref{maps1} but for a galaxy observed with the SB2 bundle. In this case the scale represents 2~kpc.}
\label{maps12}
\end{figure*}
\clearpage
\newpage

\begin{figure*}[h]
\centering
\includegraphics[width=\textwidth]{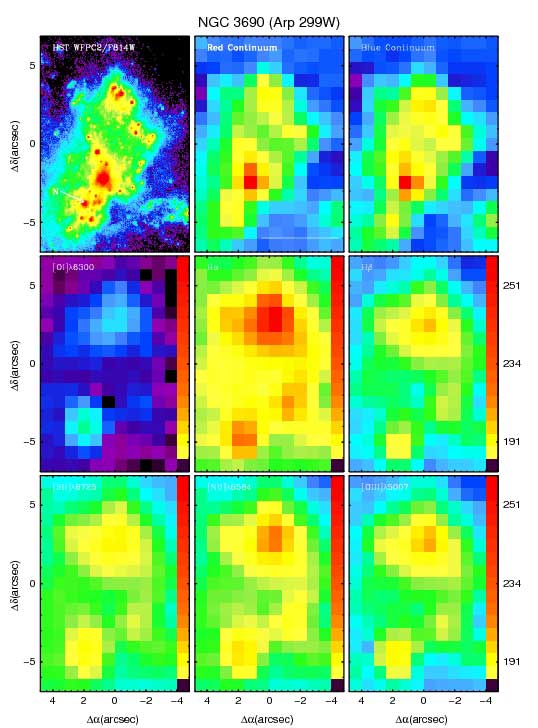} \caption{As Fig\,\ref{maps1} but for a galaxy observed with the SB2 bundle. In this case the scale represents 2~kpc.}
\label{maps13}
\end{figure*}
\clearpage
\newpage

\begin{figure*}[h]
\centering
\includegraphics[width=\textwidth]{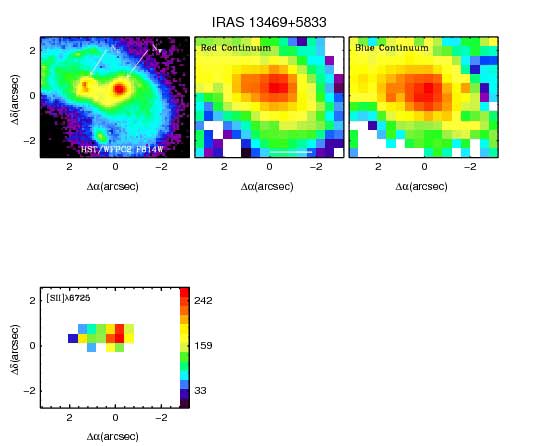} 
\caption{As Fig\,\ref{maps1} but for a galaxy observed with the SB1 bundle. In this case, the H$\alpha$+[N\,{\sc ii}] complex was strongly contaminated by a sky line. Also, the integration time was not sufficiente to achieve enough S/N ratio in the majority of the emission lines.}
\label{maps14}
\end{figure*}
\clearpage
\newpage

\begin{figure*}[h]
\centering
\includegraphics[width=\textwidth]{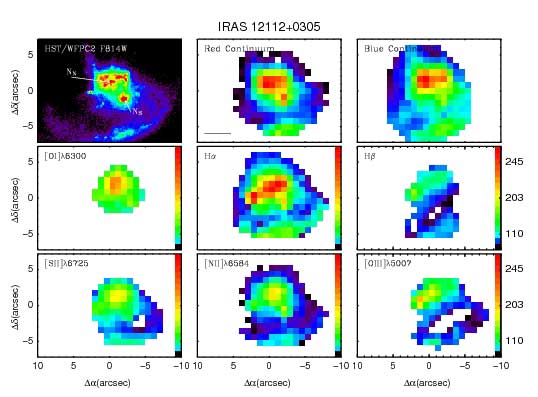} 
\caption{As Fig\,\ref{maps1} but for a galaxy where we have combined two pointings made with the SB2 bundle.}
\label{maps15}
\end{figure*}
\clearpage
\newpage

\begin{figure*}[h]
\centering
\includegraphics[width=\textwidth]{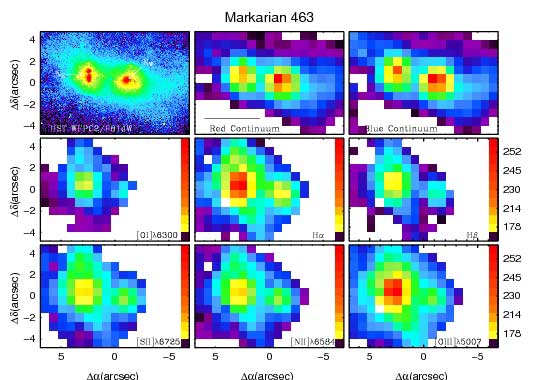} 
\caption{As Fig\,\ref{maps1} but for a galaxy observed with the SB2 bundle.}
\label{maps16}
\end{figure*}
\clearpage
\newpage

\begin{figure*}[h]
\centering
\includegraphics[width=\textwidth]{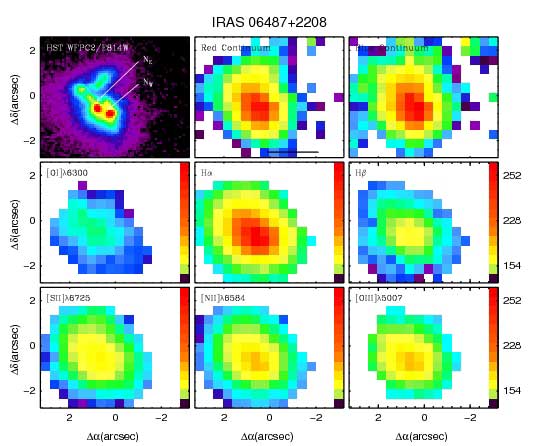} 
\caption{As Fig\,\ref{maps1} but for a galaxy observed with the SB1 bundle.}
\label{maps17}
\end{figure*}
\clearpage
\newpage

\begin{figure*}[h]
\centering
\includegraphics[width=\textwidth]{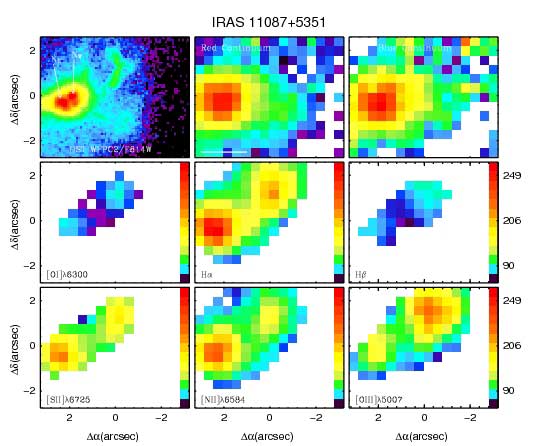} 
\caption{As Fig\,\ref{maps1} but for a galaxy observed with the SB1 bundle.}
\label{maps18}
\end{figure*}

\clearpage
\newpage

\begin{figure*}[h]
\centering
\includegraphics[width=\textwidth]{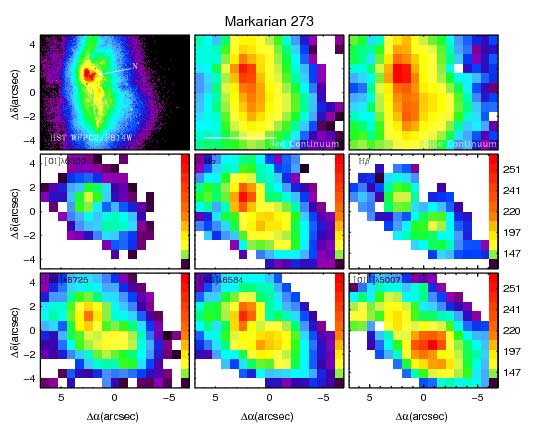} 
\caption{As Fig\,\ref{maps1} but for a galaxy observed with the SB1 bundle.}
\label{maps24}
\end{figure*}

\clearpage
\newpage

\begin{figure*}[h]
\centering
\includegraphics[width=\textwidth]{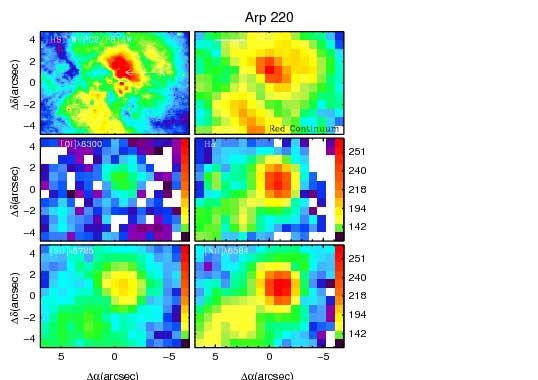} 
\caption{As Fig\,\ref{maps1} but for a galaxy observed with the SB2 bundle. In this case the scale represents 1~kpc}
\label{maps19}
\end{figure*}
\clearpage
\newpage

\begin{figure*}[h]
\centering
\includegraphics[width=\textwidth]{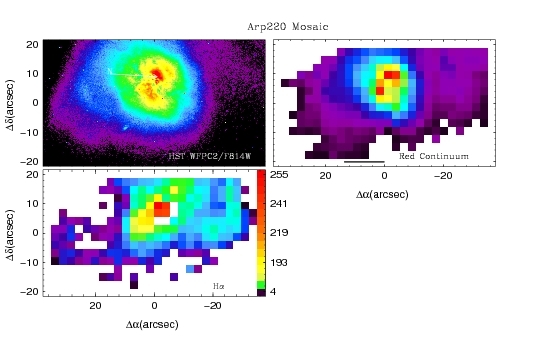} 
\caption{As Fig\,\ref{maps1} but for a mosaic obtained combining three pointings with the SB3 bundle.}
\label{maps20}
\end{figure*}

\clearpage
\newpage

\begin{figure*}[h]
\centering
\includegraphics[width=\textwidth]{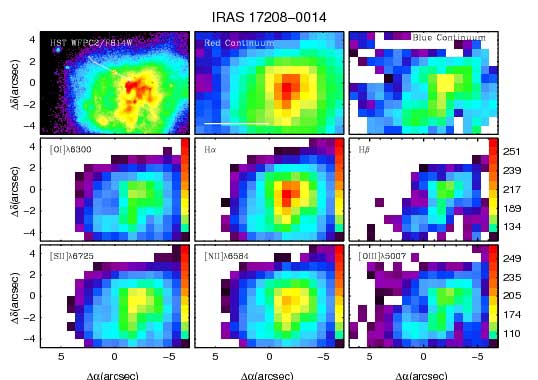} 
\caption{As Fig\,\ref{maps1} but for a galaxy observed with the SB2 bundle.}
\label{maps25}
\end{figure*}
\clearpage
\newpage

\begin{figure*}[h]
\centering
\includegraphics[width=\textwidth]{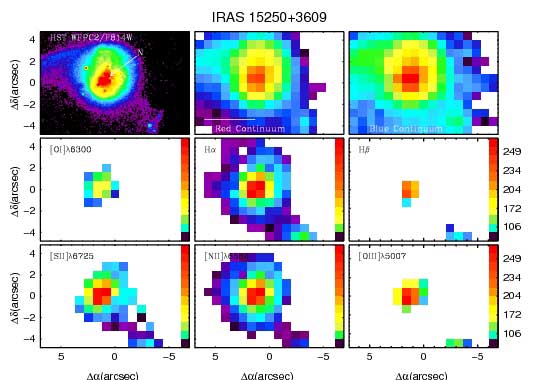} 
\caption{As Fig\,\ref{maps1} but for a galaxy observed with the SB2 bundle.}
\label{maps26}
\end{figure*}
\clearpage
\newpage

\begin{figure*}[h]
\centering
\includegraphics[width=\textwidth]{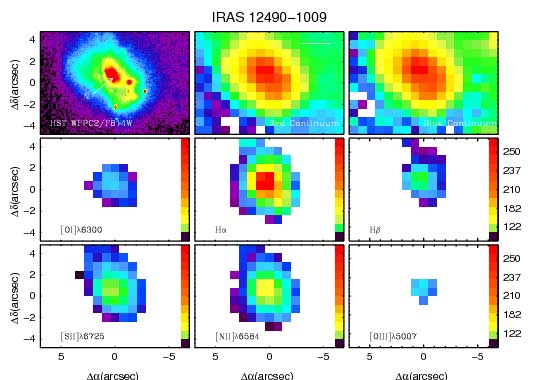} 
\caption{As Fig\,\ref{maps1} but for a galaxy observed with the SB2 bundle.}
\label{maps22}
\end{figure*}
\clearpage
\newpage

\begin{figure*}[h]
\centering
\includegraphics[width=\textwidth]{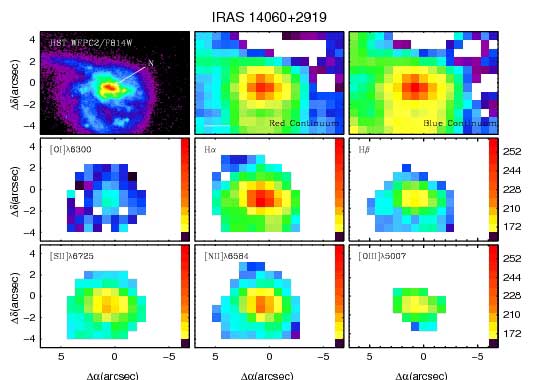} 
\caption{As Fig\,\ref{maps1} but for a galaxy observed with the SB2 bundle.}
\label{maps23}
\end{figure*}
\clearpage
\newpage

\begin{figure*}[h]
\centering
\includegraphics[width=\textwidth]{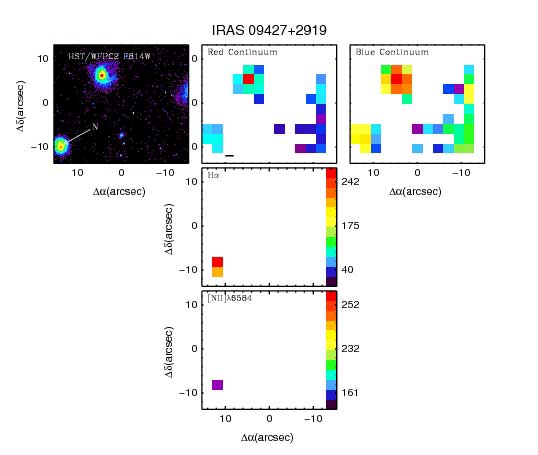} 
\caption{As Fig\,\ref{maps1} but for a galaxy observed with the SB3 bundle.}
\label{maps21}
\end{figure*}
\clearpage
\newpage

\begin{figure*}[h]
\centering
\includegraphics[width=\textwidth]{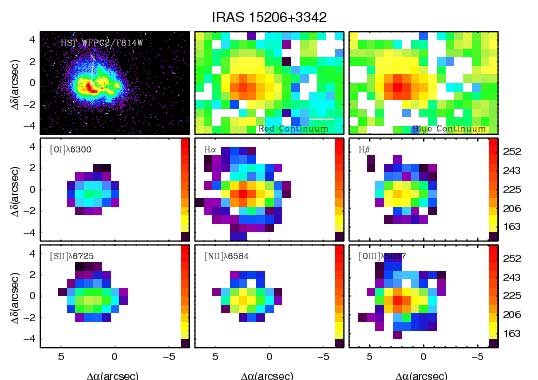} 
\caption{As Fig\,\ref{maps1} but for a galaxy observed with the SB3 bundle.}
\label{maps27}
\end{figure*}
\clearpage
\newpage

\begin{figure*}[h]
\centering
\includegraphics[width=\textwidth]{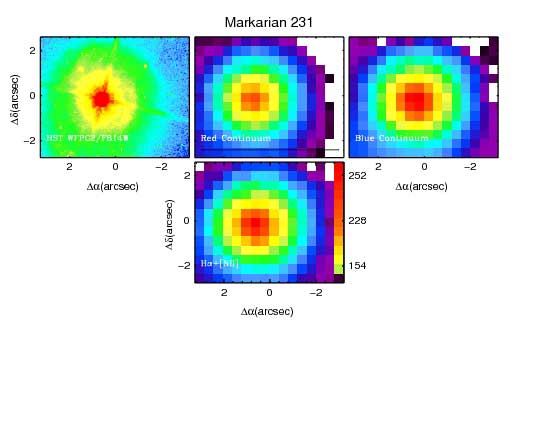} 
\caption{As Fig\,\ref{maps1} but for a galaxy observed with the SB1 bundle. The strong contamination coming from the Seyfert nucleus precludes the lines detection. Only the H$\alpha$+[N\,{\sc ii}] complex has being detected as a single structure.}
\label{maps28}
\end{figure*}

\clearpage
\newpage

\end{appendix}


\begin{thebibliography}{}

\bibitem{Adel06}
\bibitem{AAH00009}
Alonso-Herrero, A. 2009, A\&A submitted
\bibitem{b11000}
Arribas, S., Colina, L., Monreal-Ibero, A., Alfonso, J., Garc\'{\i}a-Mar\'{\i}n, M. \& Alonso-Herrero, A. 2008, A\&A, 479, 687
\bibitem{Ar2003}
Arribas, S., Colina, L., 2003, ASP Conference Proceedings, 297, 24-28
\bibitem{b11001}
Arribas, S., Colina, L. \& Borne, K. 1999, Ap\&SS, 266, 143
\bibitem{b2}
Arribas, S. et al.\,1998, Proc. SPIE, 3355, 821
\bibitem{b3}
Arribas, S., Mediavilla, E., Garc\'{\i}a-Lorenzo, B., \& del Burgo, C. 1997, ApJ, 490, 227
\bibitem{b4}
Bendo, G. J. \& Barnes, J, E. 2000, MNRAS, 316, 315
\bibitem{b5}
Bingham, R. G., Gellatly, D. W., Jenkins, C. R., \& Worswick, S. P. 1994, Proc. SPIE, 2198, 56
\bibitem{b555}
Borne, K.D., Bushouse, H., Lucas, R. A., \& Colina, L. 2000, ApJ, 529L, 77
\bibitem[2002]{Bus02}
Bushouse, H. A., Borne, K. D., Colina, L., Lucas, R. A., Rowan-Robinson, M., Baker, A. C., Clements, D. L., Lawrence, A. \& Oliver, S. 2002, ApJS, 138, 1
\bibitem{b555000}
Caputi, K. I. et al. 2007, ApJ, 660, 97
\bibitem{b55500000}
Caputi, K. I. et al. 2006a ApJ, 637, 727
\bibitem{b55500000}
Caputi, K. I. McLure, R. J., Dunlop, J. S., Cirasuolo, M. \& Schael, A. M. 2006b, MNRAS, 366, 609 
\bibitem{b5550ee00}
Clements, D. L., Sutherland, W. J., Saunders, W., Efstathiou, G. P., McMahon, R. G., Maddox, S., Lawrence, A. \& Rowan-Robinson, M. 1996, MNRAS, 279, 459
\bibitem{bb5}
Colina, L., Arribas, S. \& Monreal, A. 2005, ApJ, 621, 725
\bibitem{b7}
Colina, L., Arribas, S. \& Borne, K. D. 1999, ApJ, 527L, 13
\bibitem{Dar06000}
Darling, J. \& Giovanelli, R. 2006, AJ, 132, 2596
\bibitem{Dasy06}
Dasyra, K. M., Tacconi, L. J., Davies, R. I., Naab, T., Genzel, R., Lutz, D., Sturm, E., Baker, A. J., Veilleux, S., Sanders, D. B., \& Burkert, A. 2006, ApJ, 651, 835
\bibitem{Duc04KKK}
Elbaz, D., Cesarsky, C. J., Chanial, P., Aussel, H., Franceschini, A., Fadda, D. \& Chary, R. R. 2002, A\&A, 384, 848
\bibitem{F014444}
Farrah, D., Bernard-Salas, J., Spoon, H. W. W., Soifer, B. T., Armus, L., Brandl, B., Charmandaris, V., Desai, V., Higdon, S., Devost, D. \& Houck, J. 2007, ApJ, 667, 149
\bibitem{F01}
Farrah, D., Rowan-Robinson, M., Oliver, S., Serjeant, S., Borne, K., Lawrence,
A., Lucas, R. A., Bushouse, H., \& Colina, L. 2001, MNRAS, 326, 1333
\bibitem{F0uuuu1}
Forster Schreiber, N. M. et al. 2009, arXiv0903.1872 
\bibitem{F0uuuu144}
Forster Schreiber, N. M. et al. 2006, ApJ, 645, 1062
\bibitem{Bego2002}
Garc\'{\i}a-Lorenzo, B., Acosta-Pulido, J. A., \& Megias-Fern\'andez, E. 2002, ASP Conference Proceedings, 282, 501
\bibitem{maca2}
Garc\'{\i}a-Mar\'{\i}n , M.,Colina, L., Arribas, S., Alonso-Herrero, A. \& Mediavilla, E. 2006, \apj, 650, 850
\bibitem{maca1}
Garc\'{\i}a-Mar\'{\i}n 2007, PhD, Universidad Aut\'onoma de Madrid
\bibitem{b100tttt}
Genzel, R. et al. 2006, Nature, 442, 786
\bibitem{b100}
Genzel, R., Tacconi, L. J., Rigopoulou, D., Lutz, D., \& Tecza, M. 2001, ApJ, 563, 527
\bibitem{Genzel1998}
Genzel, R., Lutz, D., Sturm, E., Egami, E., Kunze, D., Moorwood, A. F. M., Rigopoulou, D., Spoon, H. W. W., Sternberg, A., Tacconi-Garman, L. E., et al.\,1998, ApJ, 498, 579
\bibitem{b1112}
Howarth, I. D., \& Murray, J. 1988, DIPSO A Friendly Spectrum Analysis Program (Starlink User Note 50; Chilton: Rutherford Appleton Lab.)
\bibitem{Im111223}
Imanishi, M., Dudley, C. C., Maiolino, R., Maloney, P. R., Nakagawa, T. \& Risaliti, G. 2007, ApJS, 171, 72
\bibitem{b12}
Kim, D. C., Veilleux, S. \& Sanders, D. B. 1998, ApJ, 508, 627
\bibitem{Kim95}
Kim, D.-C., Sanders, D. B., Veilleux, S., Mazzarella, J. M., \& Soifer, B. T. 1995, ApJS, 98, 129
\bibitem{Lag2005}
Lagache, G. Puget, J-L. \& Dole, H. 2005 ARA\&A, 43, 727
\bibitem{Lefl05}
Le Floc'h, E. et al. 2005, ApJ, 632, 169
\bibitem{Lefl04}
Le Floc'h, E. et al. 2004, ApJS, 154, 170
\bibitem{b129999}
Leech, K. J., Rowan-Robinson, M., Lawrence, A. \& Hughes, J. D. 1994, MNRAS, 267, 253
\bibitem{b1222}
Lonsdale, C., Farrah, D. \& Smith, H. 2006, Astrophysics Update 2, edited by John W. Mason. ISBN 3-540-30312-X. Published by Springer Verlag, Heidelberg, Germany, 285
\bibitem{b1222333}
Lonsdale, C. et al. 2004, ApJS, 154, 54
\bibitem{b122}
Low, J., Kleinmann, \& D. E. 1968, AJ, 73, 868
\bibitem{b122000}
Melnick, J. \& Mirabel, I. F. 1990, A\&A, 231L, 19
\bibitem{b14}
Mihos, J. C. \& Hernquist, L. 1996, ApJ, 464, 641
\bibitem{b13}
Mihos, J. C. 1999, Ap\&SS, 266, 195
\bibitem[1990]{Mill90}
Miller, J. S. \& Goodrich, R. W. 1990, ApJ, 355, 456
\bibitem{b155}
Monreal-Ibero, A., Colina, L., Arribas, S. \& Garc\'{\i}a-Mar\'{\i}n, M. 2007, A\&A, 472, 421
\bibitem{Monreal06}
Monreal-Ibero, A., Arribas, S. \& Colina, L. 2006, ApJ, 637, 138
\bibitem{Monreal04}
Monreal-Ibero, A. 2004, PhD Thesis, Universidad de La Laguna
\bibitem{b17000}
Moshir, M., Copan, G., Conrow, T., McCallon, H., Hacking, P.,
Gregorich, D., Rohrbach, G., Melnyk, M., Rice, W., \&  Fullmer, L. 1993, VizieR On-line Data Catalog: II/156A

\bibitem{b17}
Naab, T., Jesseit, R. \& Burkert, A. 2006, MNRAS, 372, 839

\bibitem{b17712}
Nardini, E., Risaliti, G., Salvati, M., Sani, E., Imanishi, M., Marconi, A. \& Maiolino, R. 2008, MNRAS, 385L, 130

\bibitem{b18}
Origlia L., \& Leitherer C. 2000, AJ, 119, 2018
\bibitem{b188}
P\'erez-Gonz\'alez, P. G., Rieke, G. H., Egami, E., Alonso-Herrero, A., Dole, H., Papovich, C., Blaylock, M., Jones, J., Rieke, M., Rigby, J. \& 4 coauthors 2005, ApJ, 630, 82
\bibitem{b177}
Rieke, G. H. \& Low, F. J. 1972, ApJ, 176L, 95
\bibitem{Risa06}
Risaliti, G., Maiolino, R., Marconi, A., Sani, E., Berta, S., Braito, V., Ceca, R. D., Franceschini, A., \& Salvati, M. 2006, MNRAS, 365, 303 
\bibitem{Rupk05}
Rupke, D. S., Veilleux, S. \& Sanders, D. B. 2005, ApJ, 632, 751
\bibitem{b19}
Sanders, D. B., Mazzarella, J. M., Kim, D.-C., Surace, J. A. \& Soifer, B. T. 2003, AJ, 126, 1607
\bibitem{b20}
Sanders, D. B. \& Mirabel, I. F. 1996, ARA\&A, 34, 749

\bibitem{Sanders1988b}
Sanders, D. B., Soifer, B. T., Elias, J. H., Neugebauer, G., \& Matthews, K. 1988, ApJ, 328L, 35
\bibitem{b2002}
Scoville, N. Z., Evans, A. S., Thompson, R., Rieke, M., Hines, D. C., Low, F. J., Dinshaw, N., Surace, J. A. \& Armus, L. 2000, AJ, 119, 991
\bibitem{SM1997}
Smail, Ian, Ivison, R. J., \& Blain, A. W. 1997, ApJ, 490L, 5
\bibitem{Scm03}
Schmitt, H. R., Donley, J. L., Antonucci, R. R. J., Hutchings, J. B., Kinney, A. L. \& Pringle, J. E. 2003, ApJ, 597, 768
\bibitem{So084}
Soifer, B. T., Rowan-Robinson, M., Houck, J. R., de Jong, T., Neugebauer, G., Aumann, H. H., Beichman, C. A., Boggess, N., Clegg, P. E., Emerson, J. P. \& 6 coauthors 1984, AjP, 278L, 71
\bibitem{b21}
Surace, J. A., Sanders, D. B., Vacca, W. D., Veilleux, S. \& Mazzarella, J. M. 1998, ApJ, 492, 116
\bibitem[2002]{Tac02}
Tacconi, L. J., Genzel, R., Lutz, D., Rigopoulou, D., Baker, A. J., Iserlohe, C., \& Tecza, M. 2002, ApJ, 580, 73
\bibitem{b22}
Veilleux, S., Kim, D.C., \& Sanders, D. B. 2002, ApJS, 143, 315
\bibitem{b23}
Veilleux, S., Kim, D.C., \& Sanders, D. B. 1999, ApJ, 522, 113
\bibitem{b24}
Veilleux, S., Kim, D.C., Sanders, D. B., Mazzarella, J. M., \& Soifer, B. T. 1995, ApJS, 98, 171
\bibitem{b29009}
Yan, H. et al. 2004, ApJ, 616, 63
\end{thebibliography}
\end{document}